\documentclass[12pt]{article}

\usepackage{newtxtext,newtxmath}

\usepackage[T1]{fontenc}
\usepackage{ae,aecompl}

\usepackage[numbers]{natbib}

\usepackage{hyperref}

\usepackage{graphicx}	
\usepackage{amsmath}	
\usepackage{gensymb}

\usepackage{subcaption}
\newcommand{\araa}{Annu. Rev. Astron. Astrophys.}   
\newcommand{\aj}{Astron. J.}   
\newcommand{\apj}{Astrophys. J.}   
\newcommand{\apjl}{Astrophys. J. Lett.}   
\newcommand{\apjs}{Astrophys. J. Suppl. Ser.}   
\newcommand{\aap}{Astron. Astrophys.}   
\newcommand{\mnras}{Mon. Not. R. Astron. Soc.}   
\newcommand{\nat}{Nature} 

\title{Neutron star mass estimates from gamma-ray eclipses in spider millisecond pulsar binaries}

\author{
C.~J.~Clark$^{1,2,3}$\thanks{E-mail: colin.clark@aei.mpg.de},
M.~Kerr$^{4}$,
E.~D.~Barr$^{5}$,
B.~Bhattacharyya$^{6}$,\\
R.~P.~Breton$^{3}$,
P.~Bruel$^{7}$,
F.~Camilo$^{8}$,
W.~Chen$^{5}$,
I.~Cognard$^{9,10}$,\\
H.~T.~Cromartie$^{11,12}$,
J.~Deneva$^{13}$,
V.~S.~Dhillon$^{14,15}$,
L.~Guillemot$^{9,10}$,\\
M.~R.~Kennedy$^{3,16}$,
M.~Kramer$^{5,3}$,
A.~G.~Lyne$^{3}$,
D.~Mata~S\'anchez$^{3,15,17}$,\\
L.~Nieder$^{1,2}$,
C.~Phillips$^{18}$,
S.~M.~Ransom$^{19}$,
P.~S.~Ray$^{4}$,
M.~S.~E.~Roberts$^{20}$,\\
J.~Roy$^{6}$,
D.~A.~Smith$^{21}$,
R.~Spiewak$^{3,22,23}$,
B.~W.~Stappers$^{3}$,\\
S.~Tabassum$^{24,25}$,
G.~Theureau$^{9,10,26}$,
G.~Voisin$^{3,26}$}
\date{}
\begin{document}
\maketitle
\noindent
{\footnotesize
$^{1}$ Max Planck Institute for Gravitational Physics (Albert Einstein Institute), D-30167 Hannover, Germany \\
$^{2}$ Leibniz Universit\"{a}t Hannover, D-30167 Hannover, Germany \\
$^{3}$ Jodrell Bank Centre for Astrophysics, Department of Physics and Astronomy, The University of Manchester, M13 9PL, UK\\
$^{4}$ Space Science Division, Naval Research Laboratory, Washington, DC 20375-5352, USA\\
$^{5}$ Max-Planck-Institut f\"ur Radioastronomie, Auf dem H\"ugel 69, D-53121 Bonn, Germany\\
$^{6}$ National Centre for Radio Astrophysics, Tata Institute of Fundamental Research, Pune 411 007, India\\
$^{7}$ Laboratoire Leprince-Ringuet, \'Ecole polytechnique, CNRS/IN2P3, F-91128 Palaiseau, France\\
$^{8}$ South African Radio Astronomy Observatory, Cape Town, South Africa\\
$^{9}$ Laboratoire de Physique et Chimie de l'Environnement et de l'Espace -- Universit\'e d'Orl\'eans / CNRS, F-45071 Orl\'eans Cedex 02, France\\
$^{10}$ Observatoire Radioastronomique de Nan\c{c}ay, Observatoire de Paris, Universit\'e PSL, Universit\'e d'Orl\'eans, CNRS, 18330 Nan\c{c}ay, France\\
$^{11}$ Cornell Center for Astrophysics and Planetary Science and Department of Astronomy, Cornell University, Ithaca, NY 14853, USA\\
$^{12}$ Hubble Fellowship Program Einstein Postdoctoral Fellow\\
$^{13}$ College of Science, George Mason University, Fairfax, VA 22030, resident at Naval Research Laboratory, Washington, DC 20375, USA\\
$^{14}$ Department of Physics and Astronomy, University of Sheffield, Sheffield S3 7RH, UK\\
$^{15}$ Instituto de Astrof\'{i}sica de Canarias, E-38205 La Laguna, Tenerife, Spain\\
$^{16}$ Department of Physics, University College Cork, Cork, Ireland\\
$^{17}$ Departamento de Astrof\'{i}sica, Universidad de La Laguna, E-38206 La Laguna, Tenerife, Spain\\
$^{18}$ University of Virginia, Charlottesville, VA 22904, USA\\
$^{19}$ National Radio Astronomy Observatory, 1003 Lopezville Road, Socorro, NM 87801, USA\\
$^{20}$ Eureka Scientific, Oakland, CA 94602\\
$^{21}$ Laboratoire d'Astrophysique de Bordeaux, Universit\'e de Bordeaux, CNRS, B18N, all\'ee Geoffroy Saint-Hilaire, F-33615 Pessac, France\\
$^{22}$ ARC Centre of Excellence for Gravitational Wave Discovery (OzGrav), Centre for Astrophysics and Supercomputing, Mail H29, Swinburne University of Technology, PO Box 218, Hawthorn, VIC 3122, Australia\\
$^{23}$ Centre for Astrophysics and Supercomputing, Swinburne University of Technology, PO Box 218, Hawthorn Victoria 3122, Australia\\
$^{24}$ New York University Abu Dhabi, P.O. Box 129188, Abu Dhabi, United Arab Emirates\\
$^{25}$ Department of Physics and Astronomy, West Virginia University, Morgantown, WV 26506-6315, USA\\
$^{26}$ Laboratoire Univers et Th\'eories, Observatoire de Paris, Universit\'e PSL, CNRS, Universit\'e de Paris, 92190 Meudon, France\\
\vspace{2ex}
}

\noindent

{\textbf{
    Reliable neutron star mass measurements are key to determining the equation-of-state of cold nuclear matter,     but these are rare.
    ``Black Widows'' and ``Redbacks'' are compact binaries consisting of millisecond pulsars and semi-degenerate companion stars.
    Spectroscopy of the optically bright companions can determine their radial velocities, providing inclination-dependent pulsar mass estimates.
While inclinations can be inferred from subtle features in optical light curves, such estimates may be systematically biased due to incomplete heating models and poorly-understood variability. Using data from the \textit{Fermi} Large Area Telescope, we have searched for gamma-ray eclipses from 49 spider systems, discovering significant eclipses in 7 systems, including the prototypical black widow PSR~B1957$+$20.
  Gamma-ray eclipses require direct occultation of the pulsar by the companion, and so the detection, or significant exclusion, of a gamma-ray eclipse strictly limits the binary inclination angle, providing new robust, model-independent pulsar mass constraints.
  For PSR~B1957$+$20, the eclipse implies a much lighter pulsar ($M_{\rm psr} = 1.81 \pm 0.07\,M_{\odot}$) than inferred from optical light curve modelling.
  \vspace{1ex}
}}

Since the discovery of the first ``black-widow'' pulsar, B1957$+$20, in 1988 \citep{Fruchter1988+B1957}, a sizable population of compact binary millisecond pulsar systems with semi-degenerate companion stars has emerged \citep{Roberts2012+Spiders}. These are often split into two main classes: ``black widows'' (BWs) with companion stars with masses below $M_{\rm c}\lesssim0.05\,M_{\odot}$; and ``redbacks'' (RBs) with companion masses $0.1M_{\odot} \lesssim M_{\rm c} \lesssim 0.5M_{\odot}$. The characteristic signatures of BW and RB systems are periodic disappearances of radio pulsations, often lasting for a large fraction of the orbital period. Despite being referred to as ``eclipses'', these events are too long to be caused by occultations by the companion star, but rather are explained by dispersion, scattering and absorption of radio emission by diffuse intra-binary material \citep{Stappers1996+J2051Eclipses,Polzin2020+LOFAREclipses}. This material is thought to have been ablated from the companion star's outer envelope by the intense pulsar wind. The spider nicknames come from this destructive behaviour, by analogy with arachnid species whose females have a (perhaps unfair!) reputation for killing their lighter mates.

A key motivation for finding and studying new spider pulsars is that they are one of the few types of pulsar binary system from which neutron star mass estimates can be obtained. This is because these systems have optically bright companion stars, whose radial velocities can be measured via optical spectroscopy. Dividing these by the pulsar radial velocities (measured by pulsar timing) provide binary mass ratio measurements, with which inclination-dependent mass estimates can be obtained by solving the binary mass function (see Methods - Pulsar mass constraints). Large neutron star masses have been inferred in this way from individual spider systems  \citep[e.g.][]{vanKerkwijk2011+B1957,Romani2015+J1311,Linares2018+J2215} and there are hints that spider pulsars may be systematically heavier than other species of binary neutron star \citep{Strader2019+RBSpec,Linares2020}. Several classes of theoretical neutron-star equation-of-state \citep[EoS; see][and references therein]{Ozel2016+EoS} models predict maximum masses close to $2M_{\odot}$, and so precise measurements of neutron star masses close to or above this level can have significant implications for fundamental nuclear physics.

However, spider pulsar mass estimates via radial velocity measurements depend strongly on the estimated binary inclination angle, $i$, with the inferred mass $M_{\rm psr} \propto 1/\sin^3 i$. The ability to accurately measure inclination angles for spider systems is therefore crucial if their masses are to be used to probe the nuclear EoS.

Binary inclination angles in spider systems are commonly estimated by modelling their optical light curves, which exhibit inclination-dependent features due to tidal deformation of the companion star in the pulsar's gravitational field, and heating by the pulsar. However, these models are sensitive to the exact temperature pattern on the companion star's surface, which often deviates significantly from that predicted by simple models in which the pulsar wind directly heats the inner face of the companion star, for example due to heating contributions from an intra-binary shock between the pulsar and stellar winds \citep{Romani2016+IBS}. In several RBs, this temperature pattern is even seen to vary over time \citep[e.g.][]{vanStaden2016+SpottyRB,Cho2018+VariableRBs,Clark2021+J2039}. 
Due to the $\sin^3 i$ scaling, a systematic error in the estimated inclination angle due to an incomplete heating model can lead to a large bias in the resulting pulsar mass estimate.

Millisecond pulsars (MSPs) also emit gamma-ray pulsations, as revealed by the Large Area Telescope \citep[LAT,][]{Atwood2009+LAT} onboard the \textit{Fermi Gamma-ray Space Telescope}. Gamma rays are particularly helpful in discovering and studying spider pulsars because, unlike radio waves, they are not absorbed in the diffuse intra-binary material. Partially as a result of radio searches repeatedly targeting unidentifed \textit{Fermi}-LAT sources \citep{Ray2012+PSC}, the number of known Galactic spider systems has increased tenfold since \textit{Fermi}'s launch.

The LAT data also offer a new and independent means to constrain binary inclination angles and pulsar masses in spider systems, by enabling searches for and studies of gamma-ray eclipses. \citet{Strader2016+J0427} found evidence in the \textit{Fermi}-LAT data, later confirmed by \citet{Kennedy2020+J0427} using a longer data set, for the first such gamma-ray eclipse, in a candidate accreting transitional MSP.

\begin{figure}
  \centering
	\includegraphics[width=\columnwidth]{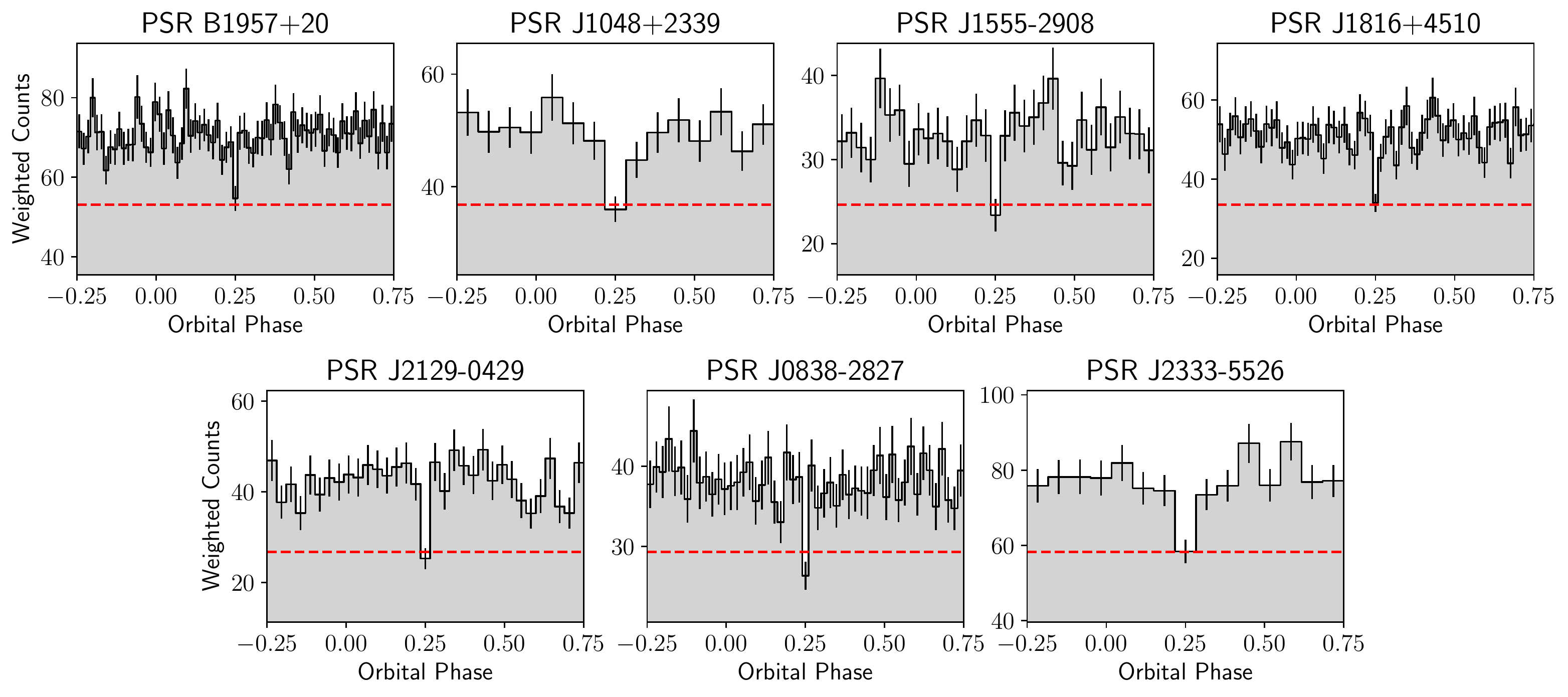}
  \caption{Gamma-ray orbital light curves of seven eclipsing spider pulsars. The red dashed line shows the estimated background level. Phase zero corresponds to the pulsar's ascending node. The phase of the pulsar's superior conjunction, where eclipses would be expected to occur, has been placed at the centre of a phase bin, and is shown at the centre of the plot for emphasis. Bin widths have been chosen to be close to the best-fitting eclipse duration. Bin heights show the sum of the photon weights in each orbital phase bin, and error bars show the corresponding 1$\sigma$ Poisson uncertainties.}
    \label{fig:eclipses}
\end{figure}

In this paper, we present a systematic search for gamma-ray eclipses from 49 confirmed/candidate spider pulsar systems. We first searched for eclipses in 42 Fermi-LAT detected confirmed spider pulsars, and found significant gamma-ray eclipses from five pulsars: PSRs~B1957$+$20, J1048$+$2339, J1555$-$2908, J1816$+$4510, and J2129$-$0429. This number of detected eclipses is consistent with the number we would expect to observe from the tested population, assuming randomly distributed orbital axes and nearly Roche-lobe filling companion stars. The orbital gamma-ray light curves for these systems are shown in Figure~\ref{fig:eclipses}, and the results of Monte-Carlo simulations used to estimate their significance (see Methods - Significance calibration via Monte-Carlo simulations) are shown in Figure~\ref{f:MonteCarlo}. Of these, the eclipse in PSR~J1555$-$2908 has the lowest significance, but still has a false-alarm probability of $4\times10^{-5}$ after accounting for the trials factor introduced by testing a range of possible eclipse widths. The most significant eclipse, from PSR~J2129$-$0429, represents a deficit of no more than 20 (weighted) photons over the 11.4 years of LAT data considered here.

For 32 of the pulsars without detected eclipses, the gamma-ray data significantly exclude otherwise feasible eclipses above a certain duration, but the faintest five systems yield no such constraints. In one of these, PSR~J0251$+$2606, there is marginal evidence for an eclipse, with a false-alarm probability of $0.002$ (see Methods - Significance calibration via Monte-Carlo simulations). Given the number of pulsars included in our search, this is around a factor of ten lower than expected for the largest outlier ($1/n_{\rm psr} \approx 0.02$). If this is indeed an eclipsing system, another seven years of accumulated \textit{Fermi}-LAT data will be required to reach the same significance as the eclipse in PSR~J1555$-$2908.

\begin{figure*}
  \includegraphics[width=\textwidth]{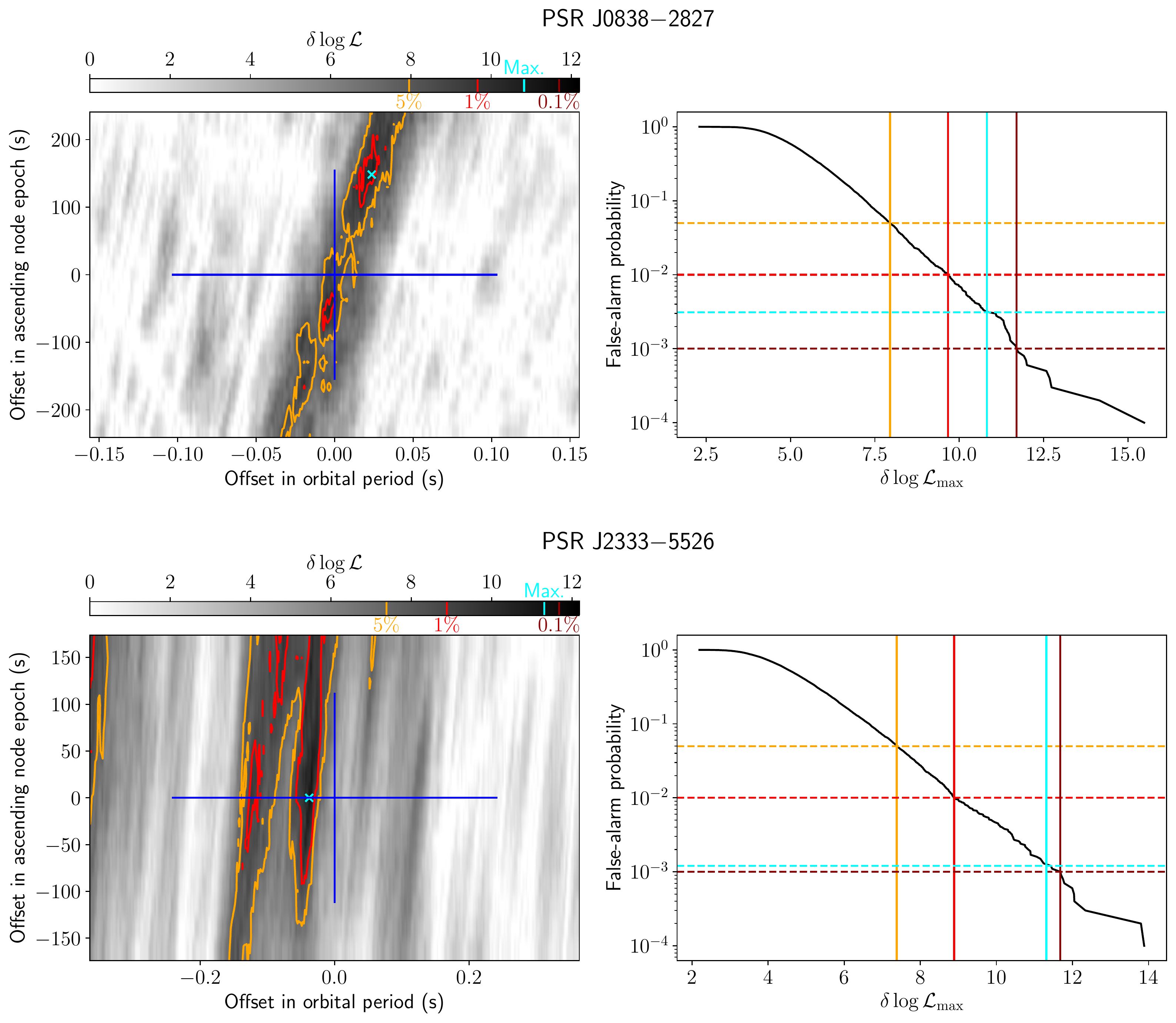}
\caption{Search results for the two candidate RB systems in which eclipses are detected. Left panels: Eclipse log-likelihoods as a function of the orbital parameters, maximised over eclipse durations. The blue crosshairs denote the $2\sigma$ ranges around the orbital ephemeris from optical observations (see references in Supplementary Material). Contour lines are drawn at log-likelihoods corresponding to false-alarm probabilities of $5\%$ (yellow) and $1\%$ (red). These levels are also marked on the colour bar, along with the log-likelihood corresponding to a false-alarm probability of $0.1\%$ (dark red), although this level is never reached. The position of the maximum likelihood is marked by a cyan cross, and the corresponding log-likelihood value is also marked in cyan on the colour bar. Right panels: the results of the Monte-Carlo simulations used to calibrate these false-alarm probabilities. The $5\%$,$1\%$ and $0.1\%$ levels are marked in the same colours used in the left panels, with the maximum log-likelihood value found in the search and corresponding false-alarm probability marked in cyan. Top panels are for PSR~J0838$-$2827, lower panels for PSR~J2333$-$5526. In all other candidate systems, the false-alarm probability for the maximum log-likelihood value found was greater than $40\%$. }  
\label{fig:candidates}
\end{figure*}

We additionally searched for gamma-ray eclipses from seven likely RB systems that were first identified from the discovery of periodic optical and/or X-ray sources within pulsar-like \textit{Fermi}-LAT sources, but which either currently lack radio or gamma-ray pulsation detections to confirm their nature, or which have only recently been detected as pulsars. These systems do not yet have precise orbital ephemerides from pulsar timing, and so a search over orbital period and phase is required, which introduces a large trials factor and therefore greatly reduces sensitivity. Nevertheless, from two of these systems, PSRs~J0838$-$2827 and J2333$-$5526, we find evidence for eclipses with trials-corrected false-alarm probabilities below $3\times10^{-3}$, which are also shown in Figure~\ref{fig:eclipses}. The search results for these two systems are shown in Figure~\ref{fig:candidates}.

We consider it unlikely that the gamma-ray eclipses are caused by the same mechanism as radio eclipses, i.e. absorption by diffuse material evaporated from the companion star. At LAT photon energies, the primary interaction between gamma rays and matter is through pair production, which for hydrogen gas has cross section $\sigma_{\gamma} = 0.03 \sigma_{\rm T}$  \citep{Petrosian1994}, where $\sigma_{\rm T}$ is the Thomson cross section. We can estimate the electron column depth in the radio eclipse region using measurements of the excess radio dispersion measure (DM). In terms of the radio DM excess, the gamma-ray optical depth is $\tau_{\gamma} = \sigma_{\gamma} \,I^{-1}\, \Delta\textrm{DM} \approx 6\times10^{-8}\, I^{-1}\, (\Delta\textrm{DM}/1\, \textrm{pc cm}^{-3})$, where $I$ is the ionisation fraction.

\citet{Polzin2020+LOFAREclipses} have studied the radio eclipses in PSRs~B1957$+$20 and J1816$+$4510. The larger DM excess was seen in J1816$+$4510 with a value of $\Delta\textrm{DM} \approx 0.1$\,pc\,cm$^{-3}$ at a phase 0.025 orbits after conjunction, corresponding to $\tau_{\gamma} \sim 6\times10^{-9} I^{-1}$. Assuming an isotropic wind, with density $\rho$ decreasing with radius $r$ from the companion star as $\rho(r) \propto r^{-2}$, the optical depth may be around 100 times higher for orbital phases half-way between conjunction and our measured eclipse egress. This is still several orders of magnitude too low to explain the observed eclipses, unless the ionisation fraction is extremely low, which seems highly unlikely given the intense environment. Similar values are obtained for B1957$+$20 and J1048$+$2339 \citep{Deneva2016+J1048}. 

Of course, this model for the companion wind is overly simplistic: spider companions have non-isotropic swept-back winds \citep{Romani2016+IBS} that often vary with time \citep{Cho2018+VariableRBs}, and the wind density profile and degree of ionisation is not yet known. If the stellar wind contributed meaningfully to the gamma-ray optical depth, we could hope to see signs of this in the \textit{Fermi}-LAT data, e.g. from gradual ingresses/egresses due to tenuous intra-binary material that increases in density towards the companion star, or time-variability in the eclipse properties. Unfortunately, the data are not sensitive to these effects, due to the very low number of expected missing photons within the relevant orbital phases. Nevertheless, for all detected eclipses, the inferred fluxes within the eclipse regions are consistent with zero, and sudden rather than gradual ingress or egresses are statistically preferred. The eclipse durations and depths also do not appear to vary over time, at least on the long timescales that we are sensitive to: eclipse widths measured from the first and second halves of the \textit{Fermi}-LAT data are consistent within their 1-sigma uncertainties, and the eclipse log-likelihoods increase approximately linearly with accumulated exposure. The observed eclipse durations are also consistent when measured in different energy bands (above or below 1 GeV).

The observed eclipses are also short enough to be caused by companion stars that fill some or all of their Roche lobes (see Table~\ref{t:eclipses}). The longest eclipse is observed in PSR~J1048$+$2339, lasting for 6--12\% of the orbital period, while the maximum eclipse duration expected for a Roche-lobe filling companion in this system is 8\%. Interestingly, optical spectroscopy has revealed emission from matter close to the L1 Lagrange point in this system, and emission lines are seen in the spectra of several other RBs, including PSR~J0838$-$2527 \citep{Halpern2017+J0838} in which we detect an eclipse with a shorter duration ($\lesssim 2\%$ of an orbit; see Supplementary Table 1). This suggests that some degree of overflowing material may be common in RB systems, but the observed eclipse durations do not currently provide evidence for gamma-ray absorption from this material. 

\begin{figure}
    \centering
	  \includegraphics[width=0.75\columnwidth]{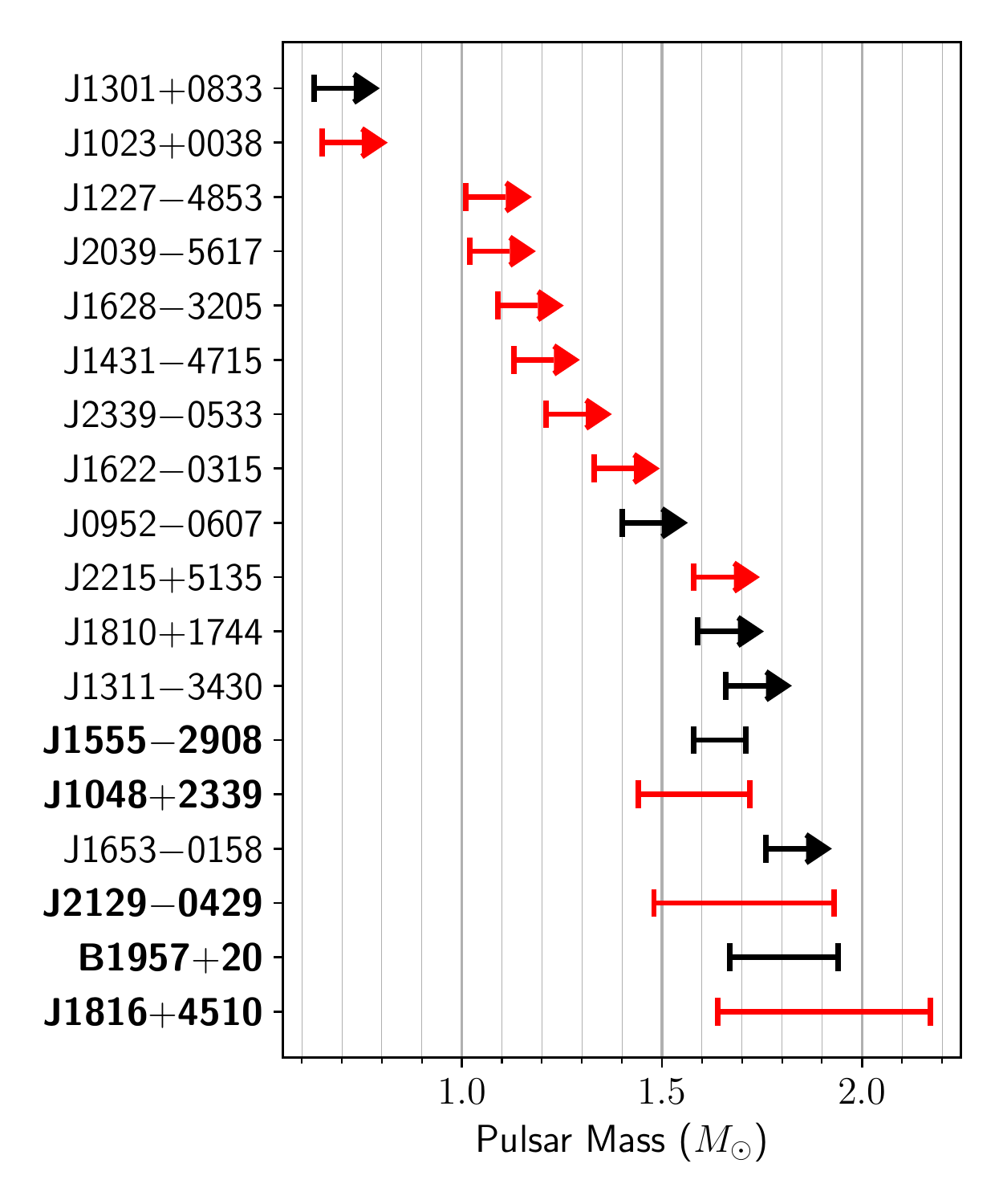}
        \caption{Neutron star mass constraints for gamma-ray detected spider MSPs, using the constraints obtained from the detection or exclusion of gamma-ray eclipses from this work. The five pulsars with detected eclipses are highlighted in bold. The two additional eclipsing systems, PSRs~J0838$-$2827 and J2333$-$5526, are excluded from this plot, as their mass ratios are not yet known from pulsar timing, and so their masses cannot yet be estimated. The colour of each point indicates the sub-class of spider system: black widows are shown in black, redbacks shown in red. For pulsars with no detected eclipses, we show lower limits on the pulsar mass, indicated by arrows with arbitrary length. }
    \label{f:mass_list}
\end{figure}

Our observations are therefore all consistent with eclipses that are solely due to occultations of the pulsar by the companion star. Under this simpler assumption, the detection of a gamma-ray eclipse and the measurement of its duration, or the significant non-detection of an eclipse, provides a robust constraint on the binary inclination. For spider systems whose companion radial velocity curves have been measured through optical spectroscopy, these inclination limits in turn constrain the pulsar masses (see Methods - Pulsar mass constraints). For eclipsing spider systems, the minimum eclipse duration provides a lower limit for the inclination, and hence an upper limit on the pulsar mass. By the same logic, we can obtain upper limits on the inclination and lower limits on the pulsar mass for systems that are not eclipsing. We list these pulsar mass constraints for eclipsing and non-eclipsing systems in Tables~\ref{t:eclipses} and \ref{t:non_eclipses}, respectively, and illustrate these results in Figure~\ref{f:mass_list}. One of the eclipsing pulsars, PSR~J1816$+$4510, has a mass upper limit larger than 2\,$M_{\odot}$. Of the non-eclipsing pulsars, the extremely compact BW binary PSR~J1653$-$0158 \citep{Nieder2020+J1653} has the largest minimum mass at $1.76\,M_{\odot}$.

The resulting inclination limits also provide crucial independent tests that can validate or falsify multiwavelength models, including optical \citep[e.g.][]{Reynolds2007+B1957LC,Draghis2019+LCs} and X-ray light curve models \citep[e.g.][]{Wadiasingh2017+IBS}, and radio and gamma-ray pulse profile models \citep[e.g.][]{Johnson2014+LCModelling}, all of which have the inclination angle as a free parameter. 

For four of the five eclipsing pulsars, our inferred inclination constraints are consistent with existing optical modelling results (or no inclination constraints from optical modelling exist in the literature, see Methods - Optical constraints for eclipsing pulsars), but this is not the case for PSR~B1957$+$20. Modelling of photometric observations of B1957$+$20 yields inclination estimates of $63^\circ \lesssim i \lesssim 67^\circ$ \citep{Reynolds2007+B1957LC,Draghis2019+LCs}. When combined with optical spectroscopy results \citep{vanKerkwijk2011+B1957}, this corresponds to an extremely high mass of $M_{\rm psr} = 2.4 \pm 0.1$\,M$_{\odot}$, higher than that of any other known neutron star. This is at odds with most EoS models which predict lower maximum neutron star masses \citep{Ozel2016+EoS}. Our detection of a gamma-ray eclipse, however, requires a much higher inclination angle, $i > 84.1^{\circ}$. This lower bound on the inclination corresponds to $M_{\rm psr} = 1.81 \pm 0.07M_{\odot}$, with the uncertainty now dominated by the radial velocity measurement and centre-of-mass correction. This mass is more consistent with the most massive neutron stars found by more robust pulsar timing studies \citep[e.g.,][and references therein]{Cromartie2019+HeavyPSR}. The nearly edge-on inclination is also more consistent with the estimates by \citet{Guillemot2012+B1957} and \citet{Johnson2014+LCModelling} obtained from modelling the radio and gamma-ray pulse profiles, implying that their assumption that the pulsar's spin becomes aligned with the orbit during recycling is correct.

How then do we interpret the light curve models of \citet{Reynolds2007+B1957LC} and \citet{Draghis2019+LCs}, which consistently estimate far lower inclinations than we find here? Compared to models with intermediate inclinations, a model with nearly edge-on inclination will predict less flux at minimum (for the same stellar temperature model) since less of the heated face is visible when the companion is in front of the pulsar. The irradiation pattern must therefore extend further around the back side of the companion star than predicted by a direct-heating model to increase the minimum flux to match the observed photometry at these orbital phases. Such excess heat could be caused by redirection of heating flux by an intra-binary shock wrapping around the companion star \citep{Romani2016+IBS}, or diffusion on the stellar surface causing heat to ``leak'' over the terminator \citep{Voisin2020+Redist}, a possibility that \citet{Reynolds2007+B1957LC} noted in their original modelling of this system. Revision of the optical modelling for this pulsar, using extended heating models such as these, will be required to resolve the tension with the inclination range inferred from our eclipse detection. All but one of the other eclipsing systems are RB, whose companions tend to have smaller temperature differences between the heated and unheated sides, and intra-binary shocks that wrap around the pulsar rather than the companion star \citep{Romani2016+IBS}, making these effects less strong for these systems. Optical observations of the remaining BW, PSR~J1555$-$2908, have been investigated with a model that takes heat diffusion into account \citep{Kennedy2022+J1555}, resulting in inclination constraints that are consistent with our eclipse detection.

We have not found a case in which previous optical modelling suggested a high inclination (and therefore a low pulsar mass) that is now ruled out by the non-detection of a gamma-ray eclipse. For PSR~J2215$+$5135, \citet{Romani2015+J2215} inferred a high inclination from optical modelling, and even found marginal evidence for a low-significance gamma-ray eclipse in the \textit{Fermi}-LAT data. However, eclipses in this system lasting longer than 0.1\% of an orbit are strongly ruled out by our longer data set, showing that this earlier hint was likely to be a chance false-alarm, and indeed more recent modelling by \citet{Linares2018+J2215} and \citet{Kandel2020+BWLCs} find lower, non-eclipsing inclinations.

Finally, we note that pulsars which are eclipsed by their companion stars also necessarily pass in front of the heated face of the companion star half an orbit later. As neutron stars are very small in size compared to their companion stars, but have intense gravitational fields, they will act as gravitational lenses, magnifying the optical flux from the companion star \citep[e.g.][]{Marsh2001+gravlens,Kailash2003+microlensing,Beskin2002+microlensing}. The exact degree of the magnification depends only on the pulsar mass and the orbital separation. The detection of this gravitational lensing would therefore provide an independent measurement of the neutron star mass. Unfortunately, the magnification due to lensing is expected to be on the order of $10^{-3}$~mag \citep{Marsh2001+gravlens}. Effects of this level can be dwarfed by both short- and long-timescale variability on the order of 0.1~mag \citep{Romani2012+J1311,Cho2018+VariableRBs}, as well as by systematic uncertainties in the underlying light curve due to incomplete heating models. Detecting the lensing effect will therefore require extremely sensitive optical photometry, stacked over several orbits to average out variability, and careful modelling to disentangle this effect from underlying heating effects. 

\section*{Methods}
\label{s:methods}
\subsection*{Gamma-ray observations}
For each system in our sample, we analysed 11.4 years of observations taken by the Large Area Telescope (LAT) onboard the \textit{Fermi Gamma-ray Space Telescope}. We selected \texttt{SOURCE}-class photons detected with reconstructed energies $50\,\textrm{MeV} < E <300\,\textrm{GeV}$, and with reconstructed directions from within a $3^\circ$ region-of-interest (RoI) around each pulsar, according to the \texttt{P8R3\_SOURCE\_V2} instrument response functions \citep{Pass8,Bruel2018+P305}. 

Sensitive unbinned-likelihood based methods for detecting eclipses \citep[e.g.][]{Kerr2019+godot} account for each photon individually, and therefore must account for the relative probability of each photon having been emitted by the target source, as opposed to by a fore/background source. This is achieved by \textit{weighting} the contribution of each photon to the relevant statistic \citep{Kerr2011+Weights}. Computing these weights requires an accurate spectral and spatial model of the emission from the target pulsar and all fore/background sources in the RoI \citep{Bruel2019+Weights}. For this, we used the 10-year incremental version (DR2) of the \textit{Fermi}-LAT Fourth Source Catalog \citep[4FGL,][]{4FGL,Ballet2020+4FGLDR2} (\url{https://fermi.gsfc.nasa.gov/ssc/data/access/lat/10yr_catalog/}) and the \texttt{gll\_iem\_v07.fits} Galactic diffuse emission and \texttt{iso\_P8R3\_SOURCE\_V3\_v1.txt} isotropic diffuse emission models to describe the diffuse background emission. The parameters of the spectra of the target pulsars were then refined such that the resulting photon weights maximize the significances of their gamma-ray pulsations, as described in \citep{Bruel2019+Weights}. These photon weights make use of the ``PSF'' event types (\url{https://fermi.gsfc.nasa.gov/ssc/data/analysis/documentation/Cicerone/Cicerone_Data/LAT_DP.html}) to benefit from the narrower point-spread function for well-reconstructed photon events. 

The required timing ephemerides for each spider pulsar were compiled as part of an upcoming third iteration of the \textit{Fermi}-LAT Pulsar Catalogue \citep{2PC}. For each pulsar, we computed the orbital phase at which each photon was emitted according to these ephemerides using the \texttt{TEMPO2} software \citep{Edwards2006+Tempo2}. The orbital ephemeris validity was verified by the presence of gamma-ray pulsations throughout the data.

\subsection*{Test statistic for eclipse detection}
To test for possible eclipses we adopted the unbinned likelihood estimation methods described in \citep{Kennedy2020+J0427} and \citep{Kerr2019+godot}. Under this model, we assume that the eclipse has sharp in/egresses, is centred on the pulsar's superior conjunction, and lasts for a fraction $\theta$ of the orbital period. Within the eclipse, we assume a constant flux level, which we parameterise with $\alpha$, the fractional flux level within the eclipse relative to the overall average flux. The increase in the log-likelihood for such an eclipse, compared to the null hypothesis of photons being uniformly distributed in orbital phase, is
\begin{equation}
    \begin{split}
    \label{e:logL}
        \delta\log\mathcal{L}(\alpha,\theta) = &\sum_{i\in\Theta}\log \left(w_i\,\alpha + 1-w_i\right)\\ &+ \sum_{i\in\bar{\Theta}}\log \left(w_i \frac{1-\alpha\theta}{1-\theta} + 1-w_i\right) \\ &-\left(\alpha\eta_{\Theta} + \frac{1-\alpha\theta}{1-\theta}\eta_{\bar{\Theta}}-1\right)\sum_i w_i,
\end{split}
\end{equation}
where $w_i$ is the photon probability weight for the $i$-th photon; $\Theta$/$\bar{\Theta}$ refer to photons with orbital phases inside/outside the eclipse, respectively; and $\eta_{\Theta}$/$\eta_{\bar{\Theta}}$ denotes the fractional exposure inside/outside the eclipse, respectively. The last term in Equation~\ref{e:logL} accounts for variations in the exposure as a function of orbital phase. We computed the exposure for each pulsar in $30$\,s time intervals over the \textit{Fermi} mission using \texttt{godot} \citep{Kerr2019+godot}, and folded these on the orbital period to compute $\eta_{\Theta}$ and $\eta_{\bar{\Theta}}$. After several years of observations, corresponding to several thousands of orbits of each pulsar system included here, the exposure is usually very evenly distributed across all orbital phases and hence exposure variations typically have very little effect on the resulting likelihood calculation; we correct for this effect nevertheless.

For each pulsar, we tested the hypothesis of a complete eclipse of the gamma-ray emission, corresponding to $\alpha = 0$, testing for $\theta \in [0,0.2)$ with fine spacing. The upper bound on this range is more than twice as large as the maximum possible eclipse duration for our studied population, assuming that companion stars do not overflow their Roche lobes. The pulsar with the smallest mass ratio in our population is PSR~J2129$-$0429, with $q\equiv M_{\rm psr}/M_{\rm c} = 3.93\pm0.06$, which would eclipse for 8.4\% of an orbit if the companion filled its Roche lobe and was observed at $i=90^{\circ}$.

For pulsars in which significant eclipses were detected we also tested alternative eclipse models with $0 < \alpha < 1$ and with curved rather than sharp ingresses and egresses. No significant log-likelihood improvements were observed.

\subsection*{Posterior photon weights}
A significant improvement in sensitivity when searching for eclipses can be obtained by incorporating into our analysis the fact that the gamma-ray emission is pulsed, i.e. that gamma-ray photons observed at pulse phases that fall within a peak in the gamma-ray pulse profile are more likely to have originated from the pulsar than from the background. To make use of this knowledge, we use the photon re-weighting method of \citet{Kerr2019+godot}, which we briefly describe here.

A photon weight, $w$, computed  as above using the spectral and spatial model of the RoI, is our best estimate for the probability of that photon having been emitted by the pulsar, before including knowledge of the pulsar rotational phase at which the photon was emitted. We can denote this as a prior probability $P(S) = w$, where $S$ denotes the binary statement that the photon was emitted by our target source. The probability for the opposite case, $B$, where the photon is emitted by a background source, is then $P(B) = 1 - w$. The re-weighting method updates our knowledge of the probability of the photon being emitted by the target source, based on the rotational phase $\phi$ at which the photon was emitted, by applying Bayes' theorem,
\begin{equation}
  P(S \,|\, \phi) = \frac{p(\phi \,|\, S) \,P(S)}{p(\phi \,|\, S) \,P(S) + p(\phi \,|\, B)\, P (B)}\,.
  \label{e:bayes_rule}
\end{equation}
Here $P(S \,|\, \phi)$ is now the posterior probability of the photon having been emitted by the pulsar, given its rotational phase; $p(\phi \,|\, S)$ is the phase distribution of photons emitted by the pulsar, i.e. the pulsar's pulse profile, which we hereafter denote as $f(\phi)$; and $p(\phi \,|\, B)$ is the phase distribution of background photons, which we can safely assume to be uniform when folding on the millisecond pulse periods of the pulsars included here, hence $p(\phi \,|\, B) = 1$. Re-writing equation \ref{e:bayes_rule} with these values gives us the re-weighting equation,
\begin{equation}
  P(S \,|\, \phi) = w^{\prime} = \frac{w f(\phi)} {w f(\phi) + 1 - w}\,.
  \label{e:posterior_weights}
\end{equation}
We hereafter refer to $w$ as the prior weights, and $w^{\prime}$ as the posterior weights. For phases within peaks of the pulse profile, where $f(\phi) > 1$, these posterior weights are always greater than the prior weights, and for phases outside of peaks, where $f(\phi) < 1$, the posterior weights are always lower. Thus, photons within pulse peaks are up-weighted, while the rest are down-weighted. When searching for eclipses, the posterior weights help to increase the detection statistic values for true eclipses by downweighting the detrimental effect of photons that by chance have high weights, and fall within the eclipse region, but whose rotational phases do not lie within a pulse peak and are therefore less likely to have been emitted by the pulsar than initially predicted by the prior weight. Similarly, photons lying outside the eclipse region but within a pulse peak, and therefore more likely to have been emitted by the pulsar, have a larger positive contribution to the eclipse log-likelihood.

To obtain the pulse profile models, $f(\phi)$, we fit a set of wrapped Gaussian functions to the prior-weighted photon phases using the maximum-likelihood method described by \citet{2PC}. The number of Gaussian functions used to model each pulse profile was chosen to minimise the Bayesian Information Criterion \citep{Schwarz1978+BIC}. 

We initially performed our search using the prior weights, but changed to using the posterior weights after finding that they significantly improved the sensitivity to eclipses. Of the four significant eclipses that were found using the posterior weights for the eclipse search (prior weights were used for PSR~J1048$+$2339 as discussed below), three were originally significantly detected with the prior weights, but the posterior weights give significantly larger log-likelihood values, with $\delta\log\mathcal{L}$ increasing by at least 2.6 for these pulsars. 
Only the eclipse from PSR~J1555$-$2908 is undetected using the prior weights, with $\delta\log\mathcal{L}$ = 3.35 compared to $\delta\log\mathcal{L}$ = 10.07 with posterior weights, likely owing to its weak overall flux but very narrow pulse peaks. 

For one pulsar in which a significant eclipse is found, PSR~J1048$+$2339, the radio timing ephemeris only covers a shorter 3-yr portion of the LAT mission, with variations in the orbital period and a low photon flux preventing generation of a full timing ephemeris using the LAT data. The radio timing ephemeris for this pulsar contains several orbital frequency derivatives to model these variations, but this ephemeris becomes highly uncertain when extrapolating outside the time interval in which it was derived. For our eclipse study, we removed these orbital frequency derivatives from the ephemeris and computed orbital phases assuming a constant orbital period. During the radio timing interval, these orbital period variations cause orbital phase shifts of up to $\sim 10^{-3}$\ orbits \citep{Deneva2016+J1048}. This is around 2\% of the duration of the eclipse detected in this system, and therefore we do not expect this additional source of uncertainty to substantially affect our results. For this pulsar, since pulsations are not observed outside the period covered by the radio ephemeris, we used the prior weights $w$ rather than the posterior weights $w^{\prime}$ when searching for eclipses. 

We include the two gamma-ray detected transitional MSPs, PSRs~J1023$+$0038 and J1227$-$4853 in our study, classifying these as RBs, as they appear to be very similar to this class when in their non-accreting state. We note that the source of their increased gamma-ray flux during the accreting states is unclear, but we assume that it also originates close to the neutron star (as indeed seems to be the case for the gamma-ray eclipsing transitional MSP candidate 4FGL~J0427.8-6704 \citealt{Strader2016+J0427,Kennedy2020+J0427}), and include data from both the accreting- and non-accreting states in our analysis. Pulsations are not detected from these pulsars in their accreting states, and so for these we again use the prior weights, rather than the posterior weights.

\subsection*{Significance calibration via Monte-Carlo simulations}
The search over the eclipse width $\theta$ introduces an unknown number of independent trials to our search. We therefore calibrated false-alarm probabilities ($P_{\rm FA}$) via Monte-Carlo analysis. For each pulsar, we took the observed set of posterior weights, randomly sampled orbital phases from a uniform distribution, computed the log-likelihood of Equation~\ref{e:logL} for the same set of $\theta$ values as used in the eclipse search, and stored the maximum value, iterating $10^{7}$ times. 

\begin{figure}
\centering
	\includegraphics[width=0.75\columnwidth]{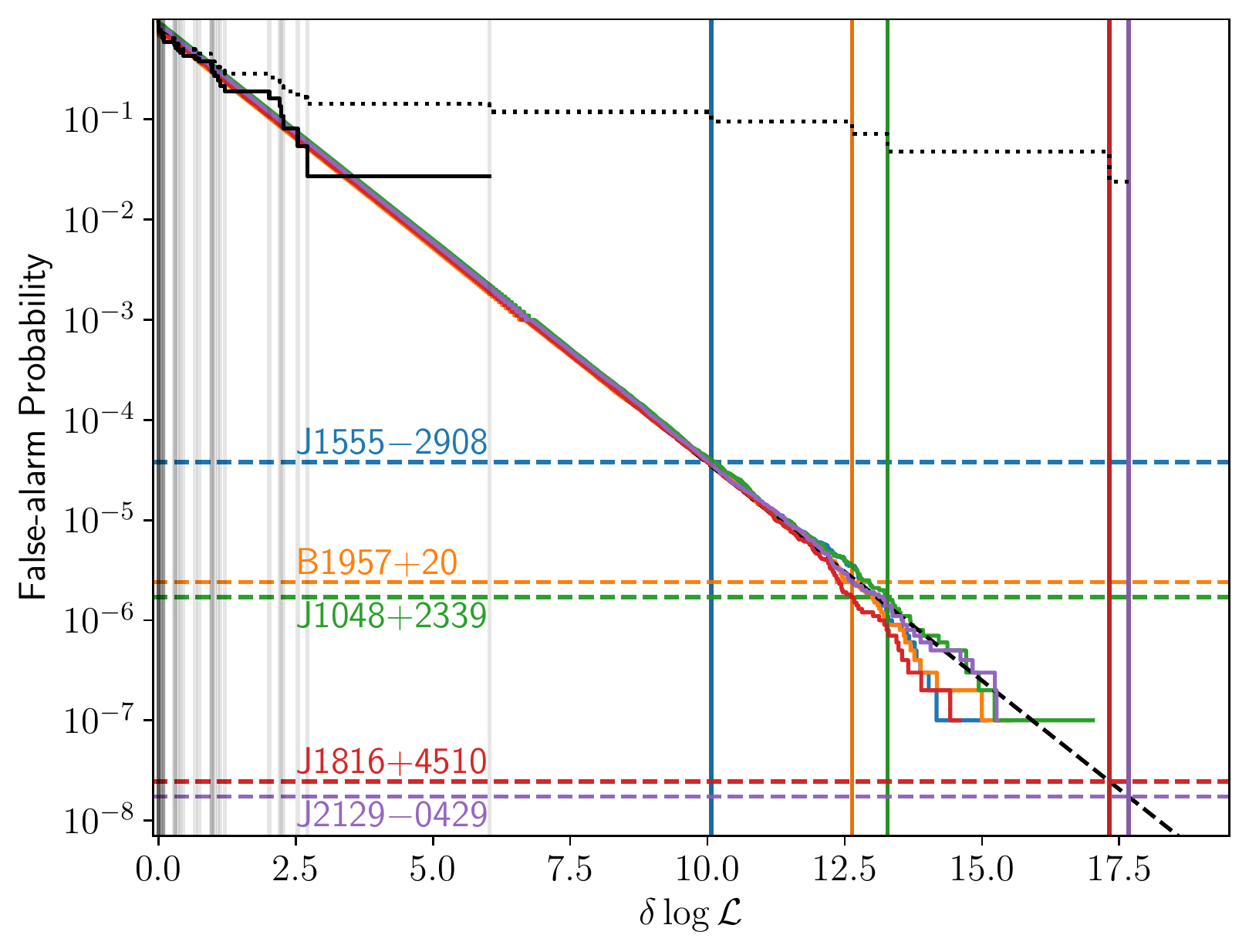}
        \caption{Results of Monte-Carlo simulations used to calibrate eclipse false-alarm probabilities. Vertical lines show the measured log-likelihood values, maximised over eclipse widths, for each pulsar. Those for pulsars with significant eclipses are marked in colour. The coloured curves show the false-alarm probability from simulations using the distribution of photon weights from each of the five eclipsing pulsars. Horizontal dashed lines show the corresponding false-alarm probability according to the Monte-Carlo calibration. The dotted and solid black curves show the empirical survival function (i.e. the fraction of pulsars which survive a given log-likelihood threshold) for the real population of spiders studied here, before and after removing the five pulsars with significant eclipses, respectively. The diagonal dashed line is an extrapolation of the fit to the simulated false-alarm probability curves used to estimate the false-alarm probabilities for the most significant eclipses.}
    \label{f:MonteCarlo}
\end{figure}

Figure~\ref{f:MonteCarlo} shows the results of the Monte-Carlo simulations that we used to calibrate the statistical significances of these eclipses. Of the five eclipsing pulsars, the eclipse in PSR J1555$-$2908 has the lowest significance, but still has a false-alarm probability $P_{\rm FA} \approx 5\times10^{-5}$.

The $\delta\log\mathcal{L}$ values observed from PSRs~J1816$+$4510 and J2129$-$0429 are larger than any obtained in our simulations. To estimate their false-alarm probabilities, we therefore performed a simple linear fit to the observed $\delta\log\mathcal{L}$ vs. $\log(P_{\rm FA})$ curves for these pulsars, and extrapolated to the observed values.

In Figure~\ref{f:MonteCarlo}, we also show the empirical survival function, i.e. the fraction of pulsars whose measured eclipse log-likelihoods would survive a given threshold. If there were no eclipses in our data set, then the set of measured log-likelihood values would be drawn from the null-hypothesis distribution and this empirical survival function curve would closely follow the simulated curves. The ratio between the empirical and simulated curves at the highest measured log-likelihood value illustrates the significance of the largest outlier, given the number of pulsars included in the sample. 

If we remove the five eclipsing systems, then the empirical survival function curve does closely follow the simulated null-hypothesis curve, and only starts to deviate for the final pulsar, PSR~J0251$+$2606, which has a false-alarm probability of around $0.2\%$. With $n_{\rm psr} = 37$ pulsars remaining in this sample, the largest outlier should have a false-alarm probability of around $1 / n_{\rm psr} = 1/37 \approx 2.7\%$. This pulsar therefore has an eclipse log-likelihood value that has an estimated false-alarm rate around ten times lower than expected for the largest outlier from our study, given the number of pulsars included. This could be viewed as marginal evidence for an eclipse, with all other measured $\delta\log\mathcal{L}$ values being consistent with the null-hypothesis. 

\subsection*{Pulsar mass constraints}
The significant detection or exclusion of a gamma-ray eclipse provides a constraint on the binary inclination angle that depends on the angular size of the companion star as seen from the pulsar. The angular size of the companion star's Roche lobe only depends on the binary mass ratio, $q$, and hence it is convenient to parameterise the size of the companion star by $q$, and its Roche-lobe filling factor $f_{\rm RL}$ (which we define as the radius of the star along the binary separation vector divided by the Roche lobe radius in the same direction). These parameters can be constrained by optical observations. The mass ratio is derived from measurements of the pulsar and companion projected radial velocity amplitudes ($K_{\rm psr}$ and $K_{\rm c}$), measured via pulsar timing and optical spectroscopy, respectively, with $q=M_{\rm psr}/M_{\rm c} = K_{\rm c}/K_{\rm psr}$. The Roche-lobe filling factor can be estimated from rotational broadening or surface gravity measurements via optical spectroscopy \citep[e.g.][]{Kaplan2013+J1816} or from the amplitude of the ``ellipsoidal'' component of an observed optical light curve \citep{Icarus}. However, this parameter is often correlated with the estimated inclination, and so previous estimates of $f_{\rm RL}$ are not necessarily consistent with new inclination constraints from an eclipse detection or exclusion. 

To compute expected eclipse durations, we generated model stars using the \texttt{Icarus} \citep{Icarus} binary modelling software. \texttt{Icarus} assumes that the surface of the star follows an equipotential contour within its Roche lobe, and therefore the simulated surface accounts for the non-spherical shape of the star due to tidal and rotational deformation. For a given binary mass ratio and Roche-lobe filling factor we can then compute the range of orbital phases at which the pulsar is eclipsed by the model star, when viewed from a given inclination. We assume that the pulsar is effectively a point-source of gamma-ray emission, since gamma-ray emission is thought to either be produced inside, or just outside, the pulsar's light cylinder \citep{Kalapotharakos2019+FP}, which is thousands of times smaller than the orbital separation in a spider binary. As we do not detect gradual in/egresses in the eclipses, and since the density profile of the outer envelope of the companion star is unknown, we assume that any line-of-sight crossing the photosphere will be fully eclipsed.

The pulsar and companion masses can be estimated, as a function of inclination, from the binary mass function,
\begin{align}
  M_{\rm psr} &= \frac{K_{\rm c}^3 P_{\rm orb} (1 + 1/q)^2}{2\pi G \sin^3 i}\,,\\
  M_{\rm c} &= \frac{K_{\rm psr}^3 P_{\rm orb} (1 + q)^2}{2\pi G \sin^3 i}\,.
\end{align}
Tables~~\ref{t:eclipses} and \ref{t:non_eclipses} list our resulting constraints on the inclination and component masses for eclipsing and non-eclipsing systems, respectively, with existing companion radial velocity measurements. When an eclipse is detected, we assume that the companion fills its Roche lobe to obtain a lower bound on the inclination, and hence a conservatively high upper bound on the pulsar and companion masses, while assuming $i=90^\circ$ provides a strict lower limit on the masses with no assumption on the filling factor. For systems without detected eclipses, we assume a low $f_{\rm RL} = 0.5$ to obtain an upper bound on the inclination, and lower bound on the component masses. This limit is based on the low filling factor for PSR~J1816$+$4510 estimated by \citet{Kaplan2013+J1816} using the surface gravity determined by optical spectroscopy. Optical models for BW and RB systems tend to have rather higher estimated filling factors \citep[e.g.][]{Draghis2019+LCs}, and so we adopt this value as a conservative estimate. 

Where possible, we take radial velocity amplitudes that have been corrected for heating effects that shift the centre-of-light away from the companion's centre-of-mass. These corrections tend to increase the inferred $K_{\rm c}$, and hence increase the pulsar mass estimate. Three RB pulsars in our list do not have published centre-of-light-corrected radial velocity amplitudes (PSRs J1431$-$4715, J1622$-$0315 and J1625$-$3205), and so the mass limits for these may be slightly underestimated. However, all three have optical light curves that suggest very little heating effect is present, so the required corrections are likely to be small for these systems.

All systems studied here that do not have radial velocity measurements are BWs (which tend to be fainter at optical wavelengths, and hence often inaccessible to spectroscopic studies). For these, we assume a typical mass ratio of $q=70$, and list the resulting inclination constraints in Supplementary Table~2. While the inclination constraints vary slowly with $q$ at values typical for BWs, the component masses do depend strongly on the assumed value of $q$, and so we do not list mass constraints here. 

\subsection*{Optical constraints for eclipsing pulsars}
Previous results from optical observations and modelling of PSR~B1957$+$20 are discussed in the main text. In the paragraphs below we discuss the existing multiwavelength observations for the other four eclipsing pulsars. Where previous works provide constraints on the Roche-lobe filling factor, we use these constraints to obtain larger (but less robust) lower limits on the inclination angle (and therefore tighter upper bounds on the pulsar masses) than are obtained by assuming $f_{\rm RL} = 1$ in the previous section. We also use these estimates to obtain upper limits on the inclination angle, rather than simply assuming $i < 90^{\circ}$ (which may imply a very low filling factor). 

Optical observations of PSR~J1048$+$2339 show significant long-term variability, with the optical maximum varying by up to a magnitude \citep{Cho2018+VariableRBs,Yap2019+J1048}. Such variability cannot yet be taken into account by precise light-curve modelling, and so no measurement of the inclination angle from optical modelling exists in the literature to date. When modelling their optical observations of this pulsar, \citet{Yap2019+J1048} fixed the inclination to $i=76^{\circ}$, the maximum value that was compatible with the lack of an observed X-ray eclipse. However, while a thermal X-ray component from the neutron star surface would indeed be eclipsed at higher inclinations, X-ray emission in RBs tends to be dominated by emission from an extended intra-binary shock, and so the lack of an X-ray eclipse does not necessarily preclude a higher inclination. \citet{Yap2019+J1048} find that $f_{\rm RL} \approx 0.85$ is compatible with multiple light curves despite long-term variability. From optical spectroscopy, \citet{MiravalZanon2021+J1048} find an observed companion radial velocity amplitude of $343.3\pm4.4$km~s$^{-1}$. Using these values, we find that the observed eclipse duration requires an inclination greater than $80.9^{\circ}$ (c.f. $80.4^{\circ}$ assuming $f_{\rm RL} = 1$ in Table~\ref{t:eclipses}). Heating corrections reduce the estimated centre-of-mass velocity to $298.7
\pm7.7$\,km\,s$^{-1}$, for a larger mass ratio (and hence lower minimum inclination of $i=80.1^{\circ}$) but a lower pulsar mass, $M_{\rm psr}\approx1.1M_{\odot}$. We use the uncorrected value of $K_{\rm c}$ in Table~\ref{fig:eclipses} to obtain a conservative bound on the pulsar mass. As noted in the main text, eclipses longer than 8\% of an orbit are consistent with the data, but would require the companion star to be significantly overflowing its Roche lobe.

PSR~J1555$-$2908 is a BW pulsar that was recently discovered by \citet{Ray2022+J1555} in a targeted radio search of a steep-spectrum radio continuum source identified within a pulsar-like gamma-ray source by \citet{Frail2018+SteepSpectrumCands}. Modelling of both optical photometry and spectroscopy by \citet{Kennedy2022+J1555}, using a model that includes the possibility of heat diffusion across the terminator, revealed the companion's projected radial velocity to be $397 \pm 2$\,km/s, and indicated a high binary inclination of $i > 75^{\circ}$, giving a maximum pulsar mass of $1.82\,M_{\odot}$. The Roche-lobe filling factor is found to be high $f_{\rm RL} > 0.93$. The duration of the gamma-ray eclipse observed here requires an inclination $83.0^{\circ} < i < 86.2^{\circ}$, for a pulsar mass $1.58\,M_{\odot} < M_{\rm psr} < 1.71\,M_{\odot}$ with the uncertainty dominated by that of the companion's radial velocity.

Optical spectroscopy of PSR~J1816$+$4510 has been modelled by \citet{Kaplan2013+J1816}. They find that this system is perhaps more similar to a white-dwarf companion than a normal RB, owing to its extremely high temperature, but due to the presence of radio eclipses that are not otherwise seen in pulsar--white-dwarf binaries we categorise it here as the latter. Detailed modelling of optical photometry to determine the inclination or Roche-lobe filling factor has not been performed, but from spectroscopic models \citet{Kaplan2013+J1816} estimated a radius that corresponds to $f_{\rm RL}\sim 0.5$, which is much smaller than observed in other RBs, motivating our use of this value as a low estimate for $f_{\rm RL}$ in Table~\ref{t:non_eclipses}. Adopting this value instead of $f_{\rm RL} = 1$ results in a higher minimum inclination of $82.6^{\circ}$ and a lower pulsar mass range of $1.68\,M_{\odot} < M_{\rm psr} < 2.11\,M_{\odot}$

For PSR~J2129$-$0429, \citet{Bellm2016+J2129}  measured a projected companion radial velocity amplitude of $K_2 = 250 \pm 4$\,km\,s$^{-1}$ (for mass ratio $q=3.93\pm0.06$), and inferred a filling factor of $f_{\rm RL} = 0.82\pm0.03$ and an inclination $i > 68^\circ$ from optical light curve modelling. With these values of $q$ and $f_{\rm RL}$, the duration of the gamma-ray eclipse requires an inclination between $76.6^{\circ} < i < 78.3^{\circ}$, consistent with the range allowed by optical modelling. This corresponds to a pulsar mass range of $1.61\,M_{\odot} < M_{\rm psr} < 1.88\,M_{\odot}$ at 95\% confidence. \citet{AlNoori2018+J2129} also found dips in the \textit{XMM-Newton} light curve for PSR~J2129$-$0429, consistent with a thermal X-ray component from the neutron star surface being eclipsed by the companion. 

\subsection*{Searching for eclipses in recently-discovered redbacks and candidates}
Seven \textit{Fermi}-LAT sources have been found to contain periodic optical and X-ray sources that are almost certainly spider binary systems \citep{Strader2014+J0523,Li2016+J0212,Linares2017+J0212,Halpern2017+J0838,Li2018+J0954,Swihart2020+J2333, Swihart2021+J0940, Li2021+J0336}. Shortly before submitting this paper, millisecond radio pulsations were detected from three of these objects (4FGL~J0838.7$-$2827, 4FGL~J0955.3$-$3949 and 4FGL~J2333.1$-$5527, hereafter PSRs~J0838$-$2827, J0955$-$3949 and J2333$-$5526, respectively) by the TRAPUM collaboration (\url{http://trapum.org/discoveries.html}), but a full timing solution is not yet available for them, and gamma-ray pulsations have not yet been detected. Pulsations have not yet been detected at any wavelength from the remaining four of these systems.  A further four similar systems \citep{Pletsch2012+J1311,Nieder2020+J1653,Ray2020+J2339,Clark2021+J2039} were initially identified in the same way, but have since been confirmed as spiders through radio or gamma-ray pulsation discoveries, and hence are already included in our search. 

To search for gamma-ray eclipses in these systems, we prepared \textit{Fermi}-LAT data sets in the same way as for the confirmed spider cases, but included 12.4 years of data. Since these data sets are not bound to the validity period of a pulsar timing ephemeris, we included this extra year of data to allow for stronger detections to partially mitigate the large trials factor (see below). For these data sets we used \texttt{gtsrcprob} to compute the photon weights, rather than optimising these to maximise the pulsation significance (since this optimisation is not possible here without a gamma-ray pulsation detection), and used the prior probability weights, since posterior weights are unavailable in the absence of pulsations.

Unlike the majority of systems studied here, where the pulsar's timing ephemeris provides precise orbital period and phase measurements over the \textit{Fermi}-LAT data set, for these systems we only have imprecise orbital phase information from optical light curves and radial velocity curves. We therefore had to additionally search small ranges of orbital phases (parameterised by each pulsar's ascending node epoch, $T_{\rm asc}$ and orbital period, $P_{\rm orb}$). We chose the search ranges to be $\pm 3\times$ the published uncertainty on these parameters. Our step sizes in each parameter were chosen such that the maximum offset on the orbital phase for each photon would be $0.001$ orbits. 

This searching introduces yet more trials in our search, reducing our sensitivity. For each of these sources, we again calibrated our search significances via Monte-Carlo simulations. We took the observed set of photon phases, assigned randomly-generated arrival times evenly distributed throughout the \textit{Fermi} mission interval, performed the search over orbital period and phase as above, and took the maximum resulting $\delta\log\mathcal{L}$, iterating 1000 times, or 10000 times if the resulting false-alarm probability was low.

Sensitivity is greatly reduced in these searches, with a false-alarm probability of $P_{\rm FA} = 0.01$ corresponding to $\delta\log\mathcal{L}\approx9$, as opposed to $\delta\log\mathcal{L}\approx4.5$ when the pulsar's precise orbital ephemeris is known. Nevertheless, there is evidence of eclipses in two systems, PSRs~J0838$-$2827 and J2333$-$5526, with $\delta\log\mathcal{L}=10.8$ and $\delta\log\mathcal{L}=11.3$, corresponding to trials-corrected false-alarm probabilities of $3\times10^{-3}$ and $1\times10^{-3}$, respectively. We show the $\delta\log\mathcal{L}$ over the searched parameter space for these two systems and the corresponding Monte Carlo calibrations in Figure~\ref{fig:candidates}.

For both PSR~J0838$-$2827 and PSR~J2333$-$5526, although significant detections are made within the searched parameter space, the eclipse likelihoods are still high towards the borders of the searched regions. We therefore searched outside this region and found higher $\delta\log\mathcal{L}$ values ($\delta\log\mathcal{L}=11.8$ and $\delta\log\mathcal{L}=14.9$, respectively), at orbital phases that are $\Delta T_{\rm asc} \approx 300$\,s and $\Delta T_{\rm asc} \approx 620$\,s later than that predicted by the ephemerides obtained from fitting the companions' radial velocity curves, corresponding to $\approx 4\sigma$ and $\approx 11\sigma$ deviations, respectively. Such offsets may be caused by the irradiation of the companion star by the pulsar, which can cause the radial velocity curve to depart slightly from a simple sinusoid, although \citet{Swihart2020+J2333} did not find any evidence for this effect in their modelling of PSR~J2333$-$5526. The recent detection of radio pulsations from these pulsars will likely clarify these tensions by providing precise orbital ephemerides.

\subsection*{Expected number of eclipsing spiders}
Although our population is small, we can also use the number of detected eclipses to probe whether or not the population of \textit{Fermi}-LAT detected spiders are viewed from randomly-distributed inclination angles. This is not necessarily expected; as many of these sources have been discovered by targeting \textit{Fermi}-LAT sources, the population may be biased towards those that are bright gamma-ray emitters, and gamma-ray emission models predict that MSPs are brightest around their rotational equator, which should in turn be aligned with the orbital plane during recycling. This would manifest in our population as a greater than expected number of eclipsing pulsars. Alternatively, if we observe a smaller number of eclipses than expected, this would be evidence that the companion stars in these systems tend to fill only a small fraction of their Roche lobes. 

Under the assumption of randomly distributed orbital axes, the binary inclination angles will be drawn from a probability distribution $p(i) = \sin(i)$, which we adopt as a prior. We restrict $i$ to $i \leq 90^{\circ}$, since systems at inclination $i$ are indistinguishable from those at $180^{\circ} - i$ as the orbital direction cannot be determined. The prior probability of a pulsar being eclipsed is therefore the integral of this prior over inclination angles greater than the minimum inclination at which a pulsar would be eclipsed, $i_{\rm ecl}$, i.e. $P(i > i_{\rm ecl}) = \cos(i_{\rm ecl})$. Assuming Roche-lobe filling companions, for typical BW and RB mass ratios ($q \equiv M_{\rm psr}/M_{\rm c}$) of $q=70$ and $q=5$, respectively, this gives a prior probability for a BW or RB being eclipsed of 10\% or 23\%, respectively. The probability of observing a certain number of eclipses from the studied population follows a binomial distribution with these success factors. From the 28 BWs and 16 RBs in our sample (including the candidates discussed in the previous section) that are bright enough for us to significantly detect or rule out an eclipse, we find 2 eclipsing BWs and 5 eclipsing RBs. The binomial probabilities for these samples are 24\% and 16\% respectively, entirely consistent with our assumptions of randomly distributed inclinations and Roche-lobe filling companions, and indeed seven eclipses is the second most likely number of eclipses to observe from the combined population. 

\section*{Data Availability}
The \textit{Fermi}-LAT data are available from the Fermi Science Support Center \url{http://fermi.gsfc.nasa.gov/ssc}. Ephemerides and folded \textit{Fermi}-LAT data sets including prior and posterior photon weights for systems with detected eclipses are available on Zenodo \url{https://doi.org/10.5281/zenodo.7133502}. Ephemerides and folded data sets for other pulsars included in this study may contain unpublished information about unrelated scientific results; these are available from the authors upon request.

\section*{Code Availability}
The Fermitools, including \texttt{gtsrcprob}, used for analysing \textit{Fermi}-LAT data, are available from \url{https://fermi.gsfc.nasa.gov/ssc/data/analysis/software/}. \texttt{TEMPO2} \citep{Edwards2006+Tempo2}, used for computing rotational and orbital photon phases is available at \url{https://bitbucket.org/psrsoft/tempo2}. \texttt{Icarus} \citep{Icarus}, used for computing expected eclipse durations to derive pulsar mass estimates, is available from \url{https://github.com/bretonr/Icarus}. \texttt{godot} \citep{Kerr2019+godot}, used for computing \textit{Fermi}-LAT exposure, is available from \url{https://github.com/kerrm/godot}. \texttt{PINT} \citep{PINT}, used for evaluating template pulse profiles to derive posterior weights, is available from \url{https://github.com/nanograv/PINT}. The scripts used to perform the eclipse searches and false-alarm calibrations are available on Zenodo \url{https://doi.org/10.5281/zenodo.7133502}. 

\section*{Acknowledgements}
C.~J.~C. would like to thank Bruce Allen for useful discussions that led to the use of posterior weights that increased the significances of the detected eclipses. We would like to thank Seth Digel, Tyrel Johnson, Melissa Pesce-Rollins, David Thompson and Zorawar Wadiasingh for carefully reviewing the manuscript on behalf of the \textit{Fermi}-LAT collaboration.

C.~J.~C., R.~P.~B, M.~R.~K., D.~M.~S. and G.~V. acknowledge support from the ERC under the European Union's Horizon 2020 research and innovation programme (grant agreement No. 715051; Spiders). This work was supported by the Max-Planck-Gesellschaft~(MPG). B.~B. acknowledges the support of the Department of Atomic Energy, Government of India, under project No. 12-R\&D-TFR-5.02-0700. Support for H.~T.~C. was provided by NASA through the NASA Hubble Fellowship Program grant \#HST-HF2-51453.001 awarded by the Space Telescope Science Institute, which is operated by the Association of Universities for Research in Astronomy, Inc., for NASA, under contract NAS5-26555. V.~S.~D. was supported by the STFC. M.~R.~K  acknowledges support from the Irish Research Council in the form of a Government of Ireland Postdoctoral Fellowship (GOIPD/2021/670: Invisible Monsters). S.~M.~R. is a CIFAR Fellow and is supported by the NSF Physics Frontiers Center award 1430284. D.~M.~S. also acknowledges the Fondo Europeo de Desarrollo Regional (FEDER) and the Canary Islands government for the financial support received in the form of a grant with number PROID2020010104.

Pulsar research at Jodrell Bank Centre for Astrophysics and access to the Lovell telescope is
supported by a consolidated grant from the UK Science and Technology Facilities Council (STFC). Work at the Naval Research Laboratory was supported by the NASA Fermi program. The MeerKAT telescope is operated by the South African Radio Astronomy Observatory, which is a facility
of the National Research Foundation, an agency of the Department of Science and Innovation.  The National Radio Astronomy Observatory is a facility of the National Science Foundation operated under cooperative agreement by Associated Universities, Inc. 

The Fermi LAT Collaboration acknowledges generous ongoing support from a number of agencies and institutes that have supported both the development and the operation of the  LAT  as  well  as  scientific  data  analysis.  These  include  the National  Aeronautics  and Space   Administration   and   the   Department   of   Energy   in the   United   States,   the Commissariat \`{a} l'Energie Atomique and the Centre National de la Recherche Scientifique /  Institut  National  de  Physique  Nucl\'{e}aire et  de  Physique  des  Particules  in  France,  the Agenzia  Spaziale  Italiana and the Istituto  Nazionale  di  Fisica  Nucleare  in  Italy,  the Ministry  of  Education,  Culture, Sports,  Science  and  Technology  (MEXT),  High  Energy Accelerator  Research Organization  (KEK)  and  Japan  Aerospace  Exploration  Agency (JAXA)  in  Japan, and  the  K.  A.  Wallenberg  Foundation,  the  Swedish  Research  Council and the Swedish National Space Board in Sweden. 

Additional   support   for   science   analysis   during   the   operations phase is gratefully acknowledged from the Istituto Nazionale di Astrofisica in Italy and the Centre National d'Etudes Spatiales  in  France. This  work  performed  in  part  under  DOE  Contract  DE-AC02-76SF00515.

Our analysis made extensive use of the \texttt{numpy} \citep{numpy}, \texttt{scipy} \citep{scipy}, \texttt{matplotlib} \citep{matplotlib}, \texttt{astropy} \citep{astropy:2013,astropy:2018} and \texttt{pycuda} \citep{pycuda} python packages.

\section*{Author Contributions Statement}
C.~J.~C. performed the eclipse search analyses, and wrote the manuscript. M.~Kerr assisted the analysis and exposure calculations. P.~B. and L.~G. produced the \textit{Fermi}-LAT data sets and photon probability weights for each pulsar. R.~P.~B, V.~S.~D., M.~R.~K., D.~M.~S. and G.~V. contributed to the interpretation of optical modelling and discussion of gravitational lensing. M.~Kerr, B.~B., R.~P.~B, V.~S.~D., M.~R.~K., L.~N., M.~S.~E.~R. and D.~M.~S. reviewed the manuscript and provided comments. All remaining authors contributed pulsar timing ephemerides required to phase-fold the \textit{Fermi}-LAT data.

\section*{Competing Interests Statement}
The authors declare no competing interests.

\clearpage

\begin{table*}
  \centering
  \scriptsize
  \caption{Constraints for pulsars with detected eclipses. $\theta^{\rm min}$ and $\theta^{\rm max}$ are the minimum and maximum eclipse durations (in orbits) at 95\% confidence. $i^{\rm min}$ is the limiting inclination at which the minimum eclipse duration would be reached for a Roche-lobe filling companion, assuming the 2$\sigma$ lower limit on the companion radial velocity amplitude $K_{\rm c}$, which is taken from the references listed in the final column, with 1$\sigma$ uncertainties quoted. $M_{\rm psr}$ and $M_{\rm c}$ give the (conservative) range of pulsar and companion masses that are allowed by the eclipse detection. The minimum masses are found by assuming $i=90^{\circ}$ and the 2$\sigma$ lower limit on $K_{\rm c}$, and require companions to substantially underfill their Roche lobes (with the exception of PSR~J1048$+$2339). The maximum masses are found by assuming the 2$\sigma$ upper limit on $K_{\rm c}$, and the minimum inclination required to produce the minimum eclipse duration with the conservative assumption of a Roche-lobe filling companion.}
  \label{t:eclipses}
  \begin{tabular}{lcccccccccc}
    \hline
 Pulsar & Class & $\delta\log\mathcal{L}$ & $P_{\rm FA}$ & $\theta^{\rm min}$ & $\theta^{\rm max}$ & $K_{\rm c}$ (km s$^{-1}$) & $i^{\rm min}$ ($^{\circ}$) & $M_{\rm psr}$ ($M_{\odot}$) & $M_{\rm c}$ ($M_{\odot}$) & Ref.\\
    \hline
    B1957$+$20 & BW & $12.63$ & $2\times10^{-6}$ & $0.007$ & $0.011$ & $353.0 \pm 4.0$ & $84.1$ & $1.67$--$1.94$ & $0.025$--$0.027$ & \citep{vanKerkwijk2011+B1957} \\
    J1048$+$2339 & RB & $13.28$ & $2\times10^{-6}$ & $0.058$ & $0.120$ & $343.3 \pm 4.4$ & $80.4$ & $1.44$--$1.72$ & $0.31$--$0.35$ & \citep{MiravalZanon2021+J1048} \\
    J1555$-$2908 & BW & $10.07$ & $4\times10^{-5}$ & $0.023$ & $0.040$ & $397.0 \pm 2.0$ & $83.1$ & $1.58$--$1.71$ & $0.057$--$0.060$ & \citep{Kennedy2022+J1555} \\
    J1816$+$4510 & RB & $17.32$ & $<1\times10^{-7}$ & $0.014$ & $0.019$ & $343.0 \pm 7.0$ & $79.0$ & $1.64$--$2.17$ & $0.18$--$0.22$ & \citep{Kaplan2013+J1816} \\
    J2129$-$0429 & RB & $17.67$ & $<1\times10^{-7}$ & $0.030$ & $0.036$ & $250.3 \pm 4.3$ & $76.3$ & $1.48$--$1.93$ & $0.39$--$0.47$ & \citep{Bellm2016+J2129} \\
    \hline
  \end{tabular}
\end{table*}

\begin{table*}
  \centering
  \scriptsize
  \caption{Constraints for pulsars without detected eclipses. Parameters are the same as in Table~\ref{t:eclipses}, but inclination upper limits $i^{\rm max}$ and mass lower limits $M^{\rm min}_{\rm psr}$ and $M^{\rm min}_{\rm c}$ are computed for the maximum eclipse duration, and assuming the companion is 50\% Roche-lobe filling, to obtain lower limits on the component masses. }
  \label{t:non_eclipses}
  \begin{tabular}{lccccccccc}
    \hline
 Pulsar & Class & $\delta\log\mathcal{L}$ & $\theta^{\rm max}$ & $K_{\rm c}$ (km/s) & $i^{\max}$ ($^{\circ}$) & $M^{\min}_{\rm psr}$ ($M_{\odot}$) & $M^{\min}_{\rm c}$ ($M_{\odot}$) & Ref.\\
    \hline
    J0952$-$0607 & BW & $0.36$ & $0.010$ & $376.1 \pm 5.1$ & $86.3$ & $1.40$ & $0.020$ & \citep{Romani2022+J0952} \\
    J1023$+$0038 & RB & $0.00$ & $0.007$ & $295.0 \pm 3.0$ & $81.9$ & $0.65$ & $0.09$ & \citep{Stringer2021+tMSPs} \\
    J1227$-$4853 & RB & $0.00$ & $0.002$ & $294.4 \pm 4.0$ & $81.1$ & $1.01$ & $0.18$ & \citep{Stringer2021+tMSPs} \\
    J1301$+$0833 & BW & $0.31$ & $0.013$ & $284.9 \pm 4.4$ & $85.7$ & $0.63$ & $0.014$ & \citep{Romani2016+J1301} \\
    J1311$-$3430 & BW & $0.00$ & $0.001$ & $633.9 \pm 5.3$ & $86.9$ & $1.66$ & $0.009$ & \citep{Romani2015+J1311} \\
    J1431$-$4715 & RB & $2.54$ & $0.066$ & $278.0 \pm 3.0$ & $90$ & $1.13$ & $0.11$ & \citep{Strader2019+RBSpec} \\
    J1622$-$0315 & RB & $0.05$ & $0.009$ & $423.0 \pm 8.0$ & $83.4$ & $1.33$ & $0.10$ & \citep{Strader2019+RBSpec} \\
    J1628$-$3205 & RB & $2.72$ & $0.009$ & $358.0 \pm 10.0$ & $82.2$ & $1.09$ & $0.14$ & \citep{Strader2019+RBSpec} \\
    J1653$-$0158 & BW & $0.74$ & $0.000$ & $700.2 \pm 7.9$ & $86.7$ & $1.76$ & $0.012$ & \citep{Nieder2020+J1653} \\
    J1810$+$1744 & BW & $0.00$ & $0.001$ & $462.3 \pm 2.2$ & $84.7$ & $1.59$ & $0.049$ & \citep{Romani2021+J1810} \\
    J2039$-$5617 & RB & $0.40$ & $0.001$ & $327.2 \pm 5.0$ & $81.7$ & $1.02$ & $0.14$ & \citep{Strader2019+RBSpec,Clark2021+J2039} \\
    J2215$+$5135 & RB & $0.00$ & $0.001$ & $412.3 \pm 5.0$ & $81.6$ & $1.58$ & $0.23$ & \citep{Linares2018+J2215} \\
    J2339$-$0533 & RB & $0.00$ & $0.000$ & $377.6 \pm 17.7$ & $81.1$ & $1.21$ & $0.24$ & \citep{Romani2011+J2339} \\
    \hline
  \end{tabular}
\end{table*}
\clearpage

\end{document}


\begin{table*}
  \centering
  \scriptsize
  \caption{Results of eclipse searches from candidate pulsars. For the two systems with false-alarm probabilities $P_{\rm FA} < 1\%$ we give the minimum and maximum eclipse durations for the orbital parameters with the maximum eclipse likelihood, and the inclination $i^{\rm min}$ at which the minimum eclipse duration is reached for a Roche-lobe filling companion star, with assumed mass ratio $q=5$. For 4FGL~J2333.1$-$5527, the eclipse durations are for the more significant detection at the revised value of $T_{\rm asc}$ found outside the initial search range (see text). }
  \label{t:candidates}
  \begin{tabular}{lcccccccc}
    \hline
 Source & Class & $\delta\log\mathcal{L}_{\rm max} $ & $P_{\rm FA}$ & $\theta_{\rm min}$ & $\theta_{\rm max}$ & $K_{\rm c}$ (km s$^{-1}$) & $i^{\rm min}$ ($^{\circ}$) & Ref. \\
    \hline
     4FGL J0212.1$+$5321 & RB & $8.22$ & $0.4$ & & & $214.1 \pm 5.0$ & & \citep{Li2016+J0212,Linares2017+J0212} \\
     4FGL J0336.0$+$7502 & BW & $4.74$ & $0.7$ & & & --- & & \citep{Li2021+J0336} \\
     4FGL J0523.3$-$2527 & RB & $4.29$ & $0.9$ & & & $190.3 \pm 1.1$ & & \citep{Strader2014+J0523} \\
    PSR J0838$-$2827 & RB & $10.82$ & $0.003$ &$0.015$ & $0.023$ & $315.0 \pm 17.0$ & $75.80$ & \citep{Halpern2017+J0838} \\
     4FGL J0940.3$-$7610 & RB & $7.19$ & $0.5$ & & & $293.2 \pm 6.0$ & & \citep{Swihart2021+J0940} \\
    PSR J0955$-$3949 & RB & $6.34$ & $0.5$ & & & $272.0 \pm 4.0$ & & \citep{Li2018+J0954} \\
    PSR J2333$-$5526 & RB & $11.31$ & $0.001$ &$0.066$ & $0.079$ & $360.0 \pm 5.0$ & $81.09$ & \citep{Swihart2020+J2333} \\
    \hline
  \end{tabular}
\end{table*}

\begin{table*}
  \centering
  \scriptsize
  \caption{Constraints for pulsars without detected eclipses, and with no published companion radial velocity amplitudes. Inclination upper limits are derived assuming $q=70$, and a 50\% Roche-lobe filling companion.}
  \label{t:non_eclipses_noopt}
  \begin{tabular}{lcccc}
    \hline
 Pulsar & Class & $\delta\log\mathcal{L}$ & $\theta^{\rm max}$ & $i^{\max}$ ($^{\circ}$) \\
    \hline
    J0023$+$0923 & BW & 0.08 & 0.007 & 85.9 \\
    J0251$+$2606 & BW & 6.03 & 0.031 & 90.0 \\
    J0312$-$0921 & BW & 2.28 & 0.007 & 86.0 \\
    J0610$-$2100 & BW & 0.08 & 0.003 & 85.8 \\
    J0636$+$5129 & BW & 0.67 & 0.062 & 90.0 \\
    J1124$-$3653 & BW & 2.02 & 0.012 & 86.3 \\
    J1446$-$4701 & BW & 0.10 & 0.004 & 85.8 \\
    J1513$-$2550 & BW & 0.00 & 0.016 & 86.9 \\
    J1544$+$4937 & BW & 0.07 & 0.007 & 85.9 \\
    J1641$+$8049 & BW & 0.09 & 0.007 & 85.9 \\
    J1745$+$1017 & BW & 1.09 & 0.007 & 86.0 \\
    J1805$+$0615 & BW & 0.30 & 0.013 & 86.5 \\
    J1833$-$3840 & BW & 0.45 & 0.044 & 90.0 \\
    J1908$+$2105 & BW & 0.98 & 0.015 & 86.8 \\
    J1946$-$5403 & BW & 1.02 & 0.005 & 85.8 \\
    J2017$-$1614 & BW & 1.13 & 0.015 & 86.7 \\
    J2047$+$1053 & BW & 0.98 & 0.011 & 86.2 \\
    J2051$-$0827 & BW & 0.29 & 0.015 & 86.7 \\
    J2052$+$1218 & BW & 0.96 & 0.040 & 90.0 \\
    J2115$+$5448 & BW & 2.24 & 0.010 & 86.2 \\
    J2214$+$3000 & BW & 0.00 & 0.000 & 85.8 \\
    J2234$+$0944 & BW & 1.21 & 0.006 & 85.9 \\
    J2241$-$5236 & BW & 0.00 & 0.001 & 85.8 \\
    J2256$-$1024 & BW & 2.21 & 0.010 & 86.1 \\
    \hline
  \end{tabular}
\end{table*}